\documentclass[runningheads, anonymous]{llncs}
\usepackage{array}
\usepackage{booktabs} 
\usepackage[T1]{fontenc}
\usepackage{wrapfig}

\usepackage{orcidlink}  
\usepackage{bbding}     

\usepackage{colortbl}
\usepackage{subcaption} 
\usepackage[utf8]{inputenc}

\usepackage{amsmath}
\usepackage{enumerate}
\usepackage{svg}
\usepackage{graphicx}

\usepackage[english]{babel}

\usepackage[linesnumbered, ruled, vlined]{algorithm2e}

\usepackage{graphicx}
\usepackage{cite}

\begin{document}
\title{NetSenseML: Network-Adaptive Compression for
Efficient Distributed Machine Learning}
\titlerunning{NetSenseML}
%


\authorrunning{Y. Wang et al.}
\institute{
}

\author{
Yisu Wang \and
Xinjiao Li  \and
Ruilong Wu \and
Huangxun Chen \and
Dirk Kutscher\textsuperscript{\Envelope}
}

\institute{
The Hong Kong University of Science and Technology (Guangzhou) \\
\email{\{ywang418,xli886,rwu408\}@connect.hkust-gz.edu.cn}, \email{huangxunchen@hkust-gz.edu.cn}, \email{dku@hkust-gz.edu.cn}
}

\maketitle              
\begin{abstract}

Training large-scale distributed machine learning models imposes considerable demands on network infrastructure, often resulting in sudden traffic spikes that lead to congestion, increased latency, and reduced throughput, which would ultimately affect convergence times and overall training performance. While gradient compression techniques are commonly employed to alleviate network load, they frequently compromise model accuracy due to the loss of gradient information. 

This paper introduces NetSenseML, a novel network adaptive distributed deep learning framework that dynamically adjusts quantization, pruning, and compression strategies in response to real-time network conditions. By actively monitoring network conditions, NetSenseML applies gradient compression only when network congestion negatively impacts convergence speed, thus effectively balancing data payload reduction and model accuracy preservation.

Our approach ensures efficient resource usage by adapting reduction techniques based on current network conditions, leading to shorter convergence times and improved training efficiency. We present the design of the NetSenseML adaptive data reduction function and experimental evaluations show that NetSenseML can improve training throughput by a factor of 1.55 to 9.84$\times$ compared to state-of-the-art compression-enabled systems for representative DDL training jobs in bandwidth-constrained conditions.


\end{abstract}

\section{Introduction}
\label{sec:intro}

Training large-scale distributed machine learning models, such as large language models \cite{touvron2023llama, gpt3}, imposes considerable demands on the network infrastructure, often resulting in sudden traffic spikes that lead to congestion, increased latency, and reduced throughput, ultimately affecting convergence times and overall training performance.  


As distributed machine learning is becoming more widely used, deployments are moving away from proprietary network technologies and isolated workloads toward shared infrastructure and multi-tenant deployments \cite{crux}. Distributed Data-Parallel (DDP) training is currently the dominant paradigm for large-scale distributed machine learning \cite{ps}. In DDP, gradient communication poses a significant challenge,
particularly due to its bursty nature: Periods of computation are followed by sudden bursts of transmission and exchange of the gradient tensor. In such shared environments, unmitigated overload can result in substantial communication volume, potential overload, and packet loss, which can significantly impact the overall training process.
Applying traditional congestion control algorithms, such as CUBIC \cite{ha2008cubic} and BBR \cite{bbr} in
protocols such as TCP \cite{tcp} and QUIC \cite{quic}, can ensure reliable transmission without overload by
adapting sending rates and thus throughput, but it would also decrease training throughput and thus
increase convergence times, since machine learning application layers can only adapt their behavior indirectly, as a response to observed transport layer throughput, if at all.

Gradient compression has emerged as a promising solution for mitigating this communication bottleneck by reducing the volume of gradient data exchanged \cite{hotnets24-gredient-compression}. Techniques such as quantization \cite{thc}, pruning \cite{han2015learning} and sparsification \cite{dgc} compress gradients to reduce the amount of data transmitted. However, existing gradient compression schemes often fail to accelerate the training process without compromising model accuracy, thereby introducing a trade-off between communication efficiency and training performance.

To address these challenges, we introduce NetSenseML, an adaptive gradient compression framework for distributed machine learning that dynamically monitors network conditions and actively responds to congestion by adjusting compression ratios. NetSenseML aims to reduce Time to Accuracy (TTA), defined as the time required to reach a target model accuracy, effectively balancing transmission efficiency, convergence speed, and overall training performance through techniques such as quantization, pruning \cite{han2015pruning, li2016pruning} and sparsification.

In the network sensing phase, NetSenseML begins by carefully monitoring gradient transmission times and estimating the available bandwidth to gauge the current state of the network, thereby adjusting the gradient compression ratio in real time based on these assessments. This adaptive approach allows NetSenseML to tailor gradient compression, in the adaptive compression phase, by appropriately applying quantization, pruning, and sparsification, aligned with the measured compression ratio and density of the gradients. This strategy ensures the optimal adjustment of the compressed gradient size to fit the prevailing network conditions and maximize bandwidth efficiency.


Our experiments demonstrate that in constrained bandwidth environments with multiple workers, NetSenseML significantly enhances the efficiency of data transmission and model training performance. Furthermore, the algorithm maintains robust data transmission and model training speeds, in multitask scenarios and under challenging bandwidth conditions, effectively balancing network utilization and training efficiency.

The main contributions of our research are as follows.
\begin{enumerate}
    \item To the best of our knowledge, NetSenseML is among the first systems to use transmission times and bandwidth estimates to detect network conditions, thereby driving the dynamic tuning of  gradient compression ratios based on perceived network conditions, improving network utilization and training efficiency.
    \item NetSenseML ensures the retention of essential gradient information, even in extreme network scenarios by adjusting compression strategies based on the compression ratio and gradient density characteristics, thus minimizing the impacts of compression on model accuracy.
    \item Rigorous testing in environments with limited bandwidth and diverse tasks shows that NetSenseML can guarantee efficient gradient transmission, stable and fast model convergence, and maintain high model accuracy, effectively minimizing the trade-offs between key performance metrics.
\end{enumerate}

\section{Scenarios for Network-adaptive DL Training}
\label{sec:motivation}

NetSenseML is designed for use in arbitrary network scenarios within Wide Area Network (WAN) environments. Training a deep neural network (DNN) involves three steps: forward propagation (FP), backward propagation (BP), and the parameter update. Distributed training addresses GPU memory constraints by dividing data across multiple GPUs, each holding a complete model copy. However, this approach requires frequent gradient synchronization across GPUs, leading to intensive communication. As models grow larger, requiring partitioning across GPUs, communication demands increase further. In such cases, the network bandwidth becomes a critical bottleneck.


We consider two WAN training scenarios:

\textbf{Scenario 1: Multi-cluster communication through a wide area network.} When computational resources are insufficient, training tasks often use wide area network connections across data centers to accelerate the process and improve resource efficiency. A typical example of this model is CloudLab\cite{duplyakin2019design}, which consists of multiple clusters distributed across various locations, providing researchers with full control over computing, networking, and storage resources.

\textbf{Scenario 2: Offloading to Public Cloud Platforms.} Multi-clusters do not always guarantee sufficient resources. For example, CloudLab\cite{duplyakin2019design} offers limited allocations per user, often requiring researchers to wait in a queue based on the usage of others. In such cases, public cloud resources serve as ideal complements to local resources. Although public cloud services are often costly, the combined use of lab resources and cloud services proves to be significantly more economical than relying solely on public cloud rentals. 

A key challenge in these environments is that the interconnection bandwidth over WANs is significantly lower than the bandwidth available within laboratory clusters or between nodes in public cloud environments. As a result, during the BP process, the computed gradients must remain in the GPU/CPU memory of the nodes, waiting until the buffered data are successfully transmitted. This delay can substantially increase the overall training time and reduce GPU utilization.

\section{Related Work}

Modern distributed machine learning frameworks, such as PyTorch DDP \cite{Pytorch}, generate a substantial volume of data packets that accumulate in communication queues, slowing down iteration speed. To address this challenge, DC2 \cite{dc2} leverages a delay monitoring component to track communication latency for each training iteration, enabling dynamic adjustment of gradient compression. Crux \cite{crux} infers network congestion by analyzing GPU utilization, allowing it to optimize the data transmission volume. Espresso \cite{wang2023hi} emphasizes the coordination between different hardware components and communication links during the compression process, aiming to optimize communication time. 

Despite existing work that estimates network conditions based on latency, GPU utilization, and bandwidth, there is a lack of effective estimation of actual network capacity.

\section{NetSenseML: Network-Aware Gradient Compression for Efficient Distributed Machine Learning}
\label{sec:design}

\begin{figure*}[t]
  \centering
    \includegraphics[scale=0.95]{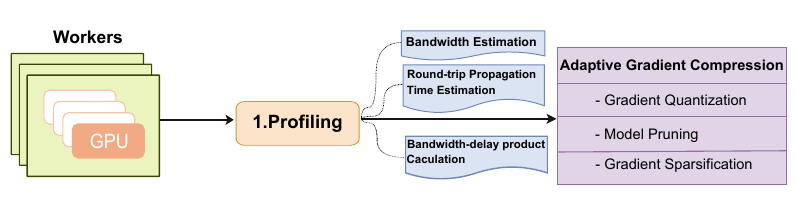}
  \caption{NetSenseML Overview}
  \label{fig:overview}
\end{figure*}

Our goal is to transmit as much gradient information as possible under limited bandwidth, unpredictable network connectivity, and constantly changing network conditions, thereby maintaining model accuracy and convergence speed. As shown in Fig. \ref{fig:overview}, the architecture includes a profiling module, which continuously monitors network status through bandwidth estimation, round-trip propagation time estimation, and bandwidth-delay product calculation, and an adaptive gradient compression module that dynamically adjusts gradient quantization and model pruning to select the optimal feasible data rate.


\subsection{Network Status Sensing and Adaptive Compression Ratio Adjustment}

\begin{wrapfigure}{r}{0.5\textwidth}
\small
  \centering
  \includegraphics[width=0.76\linewidth]{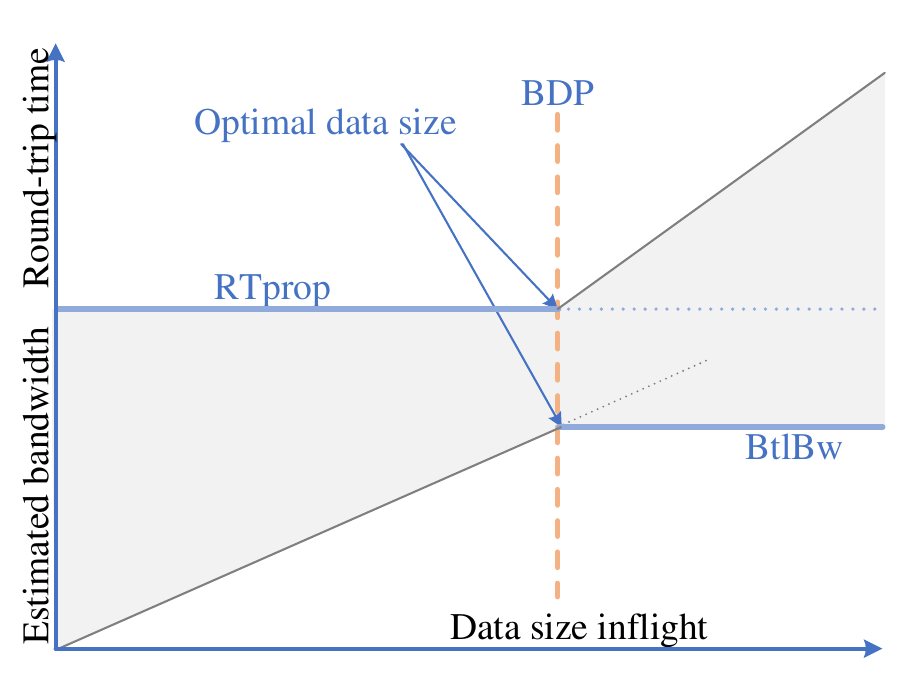}
  \caption{Network status sensing based on {\tt BtlBw} and {\tt RTprop}.}
  \label{fig:netsense}
\end{wrapfigure}

We accurately modeled the available network capacity prior to each gradient aggregation. To do this, we monitor the bottleneck bandwidth ({\tt BtlBw}) and the round-trip propagation time ({\tt RTprop}) during each gradient transmission interval to sense and analyze the state of the network as shown in 
Fig.~\ref{fig:netsense}, and adjust the compression ratio based on the bandwidth delay product ({\tt BDP}).  When the transmitted data size is less than the {\tt BDP}, the round-trip time ({\tt RTT}) remains relatively constant at its minimum value, i.e., {\tt RTprop}. 

In this phase, the network bandwidth is underutilized. However, as the size of the transmitted data exceeds the {\tt BDP}, the {\tt RTT} increases significantly, indicating a state of network congestion where the data sent surpass the network's transmission capacity, i.e., {\tt BtlBw}. Our goal is to adjust the compression ratio so that the size of the compressed gradients approximates the {\tt BDP} as closely as possible. This approach maximizes bandwidth utilization, enabling higher data throughput, which in turn enhances model accuracy and accelerates convergence speed.


\begin{algorithm*}[t]
\small
\setlength{\itemsep}{0pt} 
\caption{Network Status Sensing and Adaptive Compression Ratio Adjustment}
\label{alg:netsense}
\KwData{Transmission intervals, Data size, RTprop, Compression ratio}
\KwResult{$ratio$: Ratio used for gradient compression}
\textbf{Step 1: start-up:} \\
\Indp
    Set $\text{ratio} \gets 0.01$\;
\ForEach{step in start-up phase}{
    Quickly increase the compression ratio \;
    $\text{ratio} \gets \min(1, \text{ratio} + \beta_1)$\;
    }
\Indm

\textbf{Step 2: NetSense:} \\
\Indp
    \ForEach{gradient transmission interval $i$}{
        Measure EBB and RTT for interval $i-1$ \;
        \Indp
            $\text{EBB}_{i-1} \gets 	\frac{\text{data\_size}_{i-1} }{ \text{RTT}_{i-1}}$ \;
        \Indm
        
        \text{Update BtlBw and RTprop}\;
        \Indp
            $\text{BtlBw} \gets \text{EBB}_{Max}$\;
            $\text{RTprop} \gets \text{RTT}_{Min}$\;
        \Indm
        
        \text{Measure BDP}\;
        \Indp
            $\text{BDP} \gets \text{BtlBw} 	\times \text{RTprop}$ \;
        \Indm
        
        Adjust compression ratio \;
        \Indp
            \eIf{$data\_size > 0.9 	\times \text{BDP}$}{
            $\text{ratio} \gets \max(0.005, \text{ratio} 	\times \alpha)$ \;
        }{
            $\text{ratio} \gets \min(1, \text{ratio} + \beta_2)$ \
        }
        \Indm
}
\Indm
\end{algorithm*}

In BBR \cite{bbr}, the packet sending rate during the startup phase increases rapidly until packet loss or excessive {\tt RTT} occurs, at which point BBR detects the network's capacity limit. In NetSenseML, we initially set the gradient compression ratio to a lower value and rapidly increase this ratio during the first few steps of training until excessive {\tt RTT} is detected. 

Existing {\tt RTT}-dependent congestion control mechanisms such as MLT \cite{mlt} employ reactive strategies that activate only after substantial queue accumulation, typically when {\tt RTT} exceeds twice the propagation delay. This delayed response allows severe congestion to develop, risking packet loss that critically degrades distributed training performance.

NetSenseML introduces proactive congestion prevention through real-time bandwidth-delay product monitoring. As {\tt BDP} defines the network's maximum in-flight data capacity, approaching gradient volumes trigger immediate compression before measurable RTT increases. By maintaining transmitted data below the BDP threshold during parameter synchronization phases, our method eliminates queue buildup at intermediate nodes, sustaining optimal throughput and accelerating model convergence.


Our algorithm is comprised of two specific steps,  as shown in Algorithm. \ref{alg:netsense}:

\textbf{Step 1.} During the initial start-up phase, we set the compression ratio to 0.01. This ratio is then progressively increased over a limited number of stepsuntil the detection of packet loss or excessive {\tt RTT}.

\textbf{Step 2.} In the {\em NetSense} phase, we continuously monitor and evaluate the estimated bandwidth ({\tt EBB}) and the {\tt RTT} for each interval. These metrics are crucial for updating the values of {\tt BtlBw} and {\tt RTprop}. The compression ratio is subsequently adjusted based on the {\tt BDP}, which is calculated as the product of {\tt BtlBw} and {\tt RTprop} according Eq.~(1) to (3).
We determine the {\tt EBB} of the last interval by analyzing the size of the transmitted data and its corresponding {\tt RTT}, using the maximum observed value as the network's {\tt BtlBw}, $\text{BtlBw} = \max(\text{EBB})$, and the minimum recorded {\tt RTT} as the {\tt RTprop}, $\text{RTprop} = \min(\text{RTT})$, under conditions free from network congestion. For data size $\text{data\_size}_i$ and transmission time $\text{RTT}_i$ of interval $i$, the {\tt EBB} is calculated by Eq.~(\ref{EBB}),
\begin{equation}
\label{EBB}
\text{EBB}_i(\text{data\_size}_i, \text{RTT}_i) = \frac{\text{data\_size}_i}{\text{RTT}_i}.
\end{equation}
After updating the value of {\tt BtlBw} and {\tt RTprop}, {\tt BDP} is then estimated using Eq.~(\ref{BDP}), based on $\text{RTprop}$ and $\text{BtlBw}$ prior to each data transmission,
\begin{equation}
\label{BDP}
\text{BDP} = \text{BtlBw} \times \text{RTprop}.
\end{equation}
Consequently, the compression ratio is adjusted in accordance with {\tt BDP}, as defined in Eq.~(\ref{Adpative_Ratio}),
\begin{equation}
\label{Adpative_Ratio}
	\text{ratio}=\left\{
	\begin{aligned}
		&\max(0.005, \text{ratio} \times \alpha), &  \text{data\_size}  >   0.9 \times \text{BDP} \\
		&\min(1,     \text{ratio} + \beta_2),     &  \text{data\_size} \leq 0.9 \times \text{BDP}
	\end{aligned}\right.
\end{equation}
In our experiment, $\alpha$ is $0.5$, $\beta_2$ is $0.01$.

\subsection{Adaptive Compression Based on Quantization, Pruning, and Sparsification}

\begin{wrapfigure}{r}{0.6\textwidth}
\small
  \centering
  \includegraphics[width=1\linewidth]{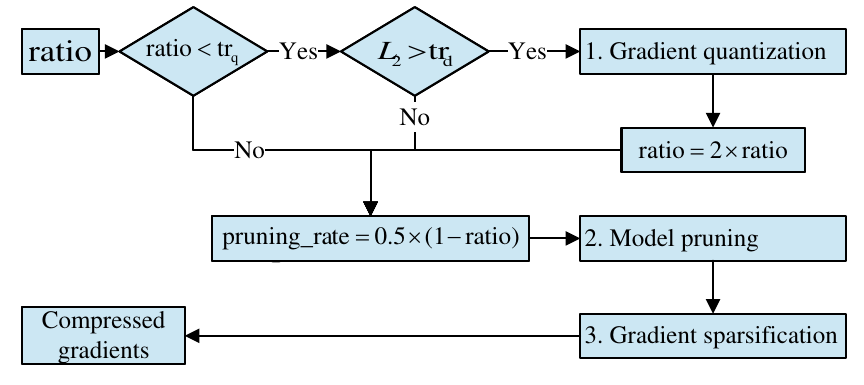}
  \caption{Adaptive quantization based on $L_2$ norm.}
  \label{fig:quantization}
\end{wrapfigure}
To enhance the precision and accelerate the convergence of the model, we have developed an adaptive compression method that incorporates quantization, pruning, and sparsification, as shown in Algorithm .~\ref{alg:adaptiveCompression}. This method optimizes gradient transmission under challenging network conditions through quantization, boosts model generalizability via pruning, and maintains critical gradient information through sparsification.
Our method consists of the following steps:
\begin{algorithm}[t]
\small
\setlength{\itemsep}{0pt} 
\caption{NetSenseCompression: quantization, pruning, and sparsification}
\label{alg:adaptiveCompression}
\KwData{
    Pre-trained model: The initial neural network model, \\
    $grads$: Gradients for compression,\\
    $\text{ratio}$: Ratio used for gradient compression, \\
    $\text{tr}_q$: Quantization check threshold,\\
    $\text{tr}_d$: Gradient density threshold,\\
}
\KwResult{$Com\_grads$: Compressed gradients}
\BlankLine
\textbf{Step 1: Adaptive Quantization}:\\
\Indp
    \If{$ \text{ratio}<\text{tr}_q$}{
    $L_2 \gets \text{calculate\_L2\_norm}(grads)$\;
    \If{$L_2 > tr_d$}{
        $grads \gets \text{quantize}(grads)$\;
        $\text{ratio} \gets 2 \times \text{ratio}$\;
    }
    }
\Indm
\textbf{Step 2: Model Pruning}:\\
\Indp
    \text{Get pruning rate}\;
    \Indp
        $\text{pruning\_rate} \gets 0.5 \times (1 - \text{ratio}) $\;
    \Indm
    
    $\text{pruned\_grads} \gets \text{prune\_weights}(\text{pruning\_rate})$\;
    \text{Set pruned weights' gradients to zero}\;
    \Indp
    $grads[\text{pruned\_grads}] \gets 0$\;
    \Indm
\Indm

\textbf{Step 3: Gradient Sparsification}:\\
\Indp
    $Com\_grads \gets \text{top\_k}(grads, \text{ratio})$\;
    \text{Transmit the sparsified gradient over the network}\;
\Indm
\end{algorithm}

\textbf{Step 1. Adaptive quantization.} Adaptive quantization is initiated when the compression ratio falls below a pre-defined threshold, $\text{tr}_q$, as shown in Fig.~\ref{fig:quantization}. The gradient's $L_2$ norm is computed as follows: \begin{itemize} \item If $L_2 > \text{tr}_d$, indicating substantial informational content, quantization is applied to preserve non-zero gradient transmission. This is essential for maintaining the gradient's integrity under stringent network conditions. \item To accommodate the halved data size resulting from the reduction of gradient representation from 32-bit to 16-bit floating points, the compression ratio is adjusted to $\text{ratio} = 2 \times \text{ratio}$. This efficient compression aids in optimizing bandwidth use while ensuring data integrity and model performance. \end{itemize}

\textbf{Step 2: Model Pruning.} The pruning rate is set using $\text{ratio}_p = 0.5 \times (1 - \text{ratio})$, which targets parameters with smaller weights for pruning: 

This process involves setting the gradients of the selected parameters to zero, which not only reduces the load during network transmission by eliminating these gradients from the next sparsification step but also minimizes the disproportionate influence of small weights with large gradients on model training. Such selective pruning is crucial for enhancing the model's generalizability and overall performance.

The parameters affected by pruning are not permanently removed but are only excluded from gradient transmission. This allows these pruned parameters the potential to be gradually reactivated in subsequent training iterations based on new data, thereby maintaining the adaptability and depth of the model’s learning capability.

\textbf{Step 3. Gradient sparsification.} 
We employ the compression ratio as the basis for the sparsification ratio, performing TopK sparsification \cite{topk}
to eliminate gradients with minimal absolute values, and accumulate the local filtered gradients for further aggregation and transmission. This process is calibrated to ensure that the decision to filter gradients is influenced equally by the model's pruning strategy and the inherent magnitude of the gradients.

This structured approach not only streamlines the handling of data in constrained network environments but also strategically manipulates the model parameters to promote faster convergence and enhance the general robustness of the model. 








\section{Implementation and Evalution}
\label{sec:eval}

\subsection{Setup and workloads}
We built our testbed using the ESXi virtualization platform with a server equipped with 8 A40 GPUs and an Intel Xeon Platinum 8358P processor with 64 CPU cores. 

We developed the NetSenseML prototype on top of the PyTorch distributed framework \cite{Pytorch}, implementing  communication using NCCL \cite{nccl} over TCP for
gradient aggregation. In our implementation, we utilized PyTorch's Distributed Data Parallel communication hook to override the default
{\tt allreduce} operation, allowing fine-grained control over how gradients
are communicated across workers.

\textbf{Workloads:} We evaluated our NetSenseML prototype by training popular computer vision models on an eight-worker testbed. The models and datasets used in our evaluation included ResNet18 \cite{resnet18} and VGG16 \cite{vgg}, both trained on the CIFAR-100 dataset. We set the per-GPU batch size to 32.

\begin{wrapfigure}{r}{0.4\textwidth}
\centering
    \includegraphics[width=1\linewidth]{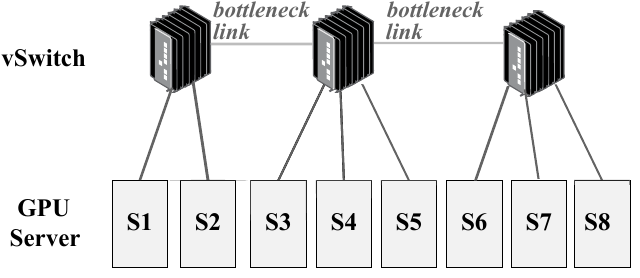}
    \caption{Evaluation topology.}
    \label{fig:topology}
\end{wrapfigure}

\textbf{Metrics:} We define time-to-accuracy ({\tt TTA}) as the training time required to reach a target validation accuracy. Additionally, we present the training throughput (measured as images per second, referred to as samples per second) across all training tasks under different scenarios and varying bandwidth conditions. We define convergence time as the time required for the model’s accuracy to stabilize at a target threshold, indicating the model has fully converged.

\subsection{Evaluation Scenarios}

We conducted experiments under three different scenarios using the topology shown in Fig.~\ref{fig:topology}, creating network performance bottlenecks by adjusting the link bandwidth of two connections to the switch.

\textbf{Scenario 1: Bandwidth constrained but stable network.} We ran several tests using different static bottleneck bandwidths, ranging from 200 Mbps to 10 Gbps, by adjusting the link bandwidth between the switch and other nodes. The bandwidth selection was based on the proportion of the maximum bandwidth required by a training model to avoid latency. These tests, which measured time-to-accuracy, convergence time, and training throughput, showed that NetSenseML can fully utilize the available bandwidth (when TopK and AllReduce cannot), leading to faster convergence. 


\textbf{Scenario 2: Degrading Network Conditions.} We gradually reduced all communication bandwidths during the experiment to demonstrate the adaptability of our approach. Training throughput was compared to show that as network conditions deteriorate, NetSenseML can achieve higher throughput by reducing gradient transmission, whereas TopK and AllReduce maintain a constant data volume, leading to network congestion and reduced training throughput.

\textbf{Scenario 3: Fluctuating Bandwidth with Competing Network Traffic.} We simulated a cloud service environment where multiple virtual machines share the same physical host's network resources. Specifically, we ran multiple iperf3 \cite{iperf} processes in parallel between nodes, periodically sending traffic to test the available bandwidth of the current network links. This generated network interference that competed for bandwidth with the training task. By dynamically adjusting the link bandwidth between switches and nodes, we evaluated the robustness of our approach. 



\subsection{End-to-End Performance}

For assessing end-to-end performance, we measured 1) {\em time-to-accuracy}, 2) {\em training throughput and communication efficiency}, 3) {\em dynamic training throughput in degrading network conditions}, and 4) {\em dynamic training throughput in fluctuating network conditions}.

\textbf{Time-to-accuracy.} We evaluated NetSenseML against TopK compression (0.1 rate) and AllReduce for ResNet18 and VGG16 models in bandwidth constrained environments (200 Mbps, 500 Mbps, and 800 Mbps). The ResNet18 model size is 46.2MB, making AllReduce impractical without at least 500 Mbps to prevent congestion. Therefore, we tested under bandwidths of 200 Mbps, 500 Mbps, and 800 Mbps.

Under extremely low bandwidth conditions, NetSenseML demonstrates superior convergence speed compared to both AllReduce and TopK. We use the point at which NetSenseML achieves its best test accuracy as the benchmark and terminate the training of AllReduce and TopK at that point. 

\begin{figure*}[t]
  \centering
  \begin{subfigure}[b]{0.32\textwidth}
    \centering
    \includegraphics[scale=0.11]{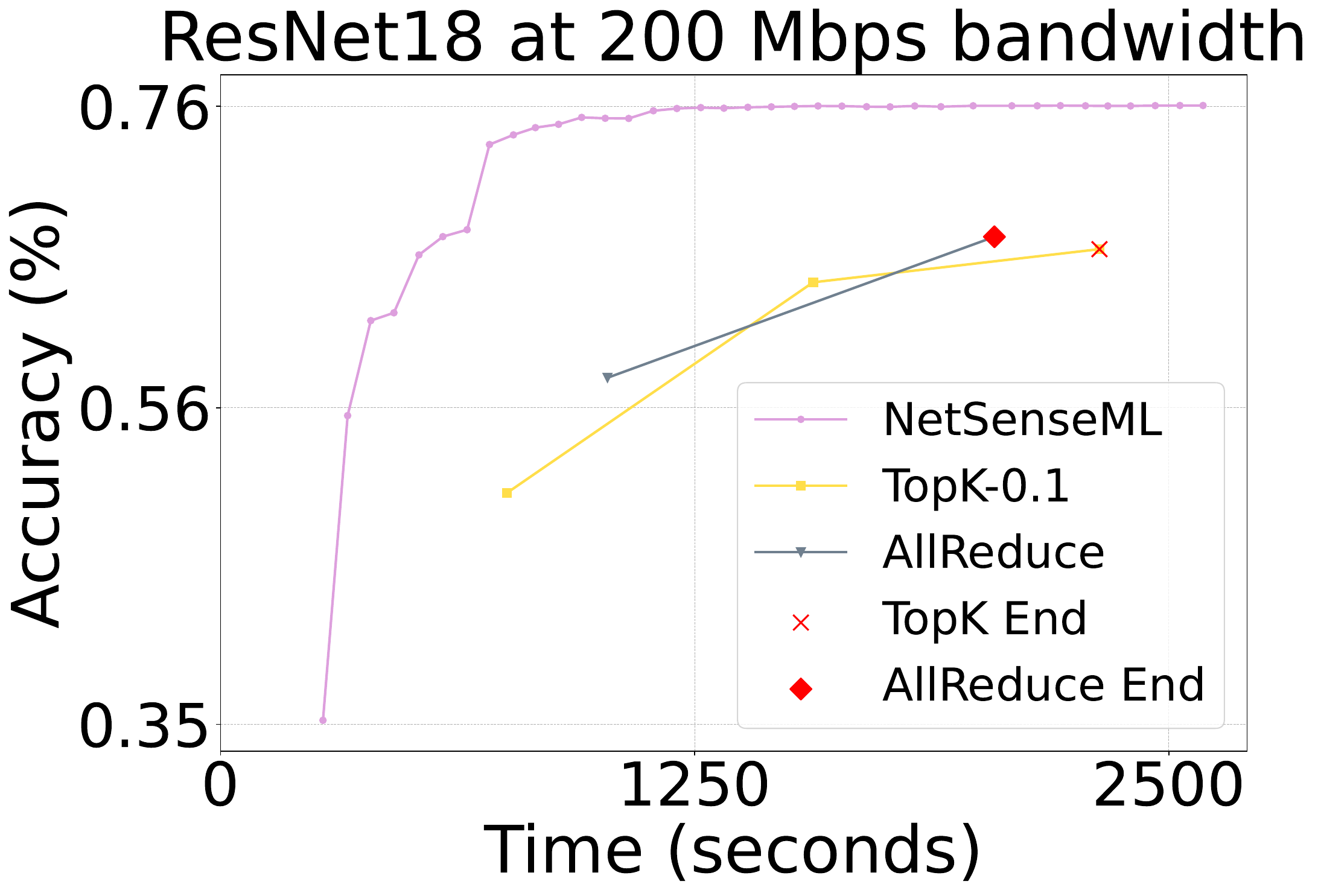}
  \end{subfigure}
  \begin{subfigure}[b]{0.32\textwidth}
    \centering
    \includegraphics[scale=0.11]{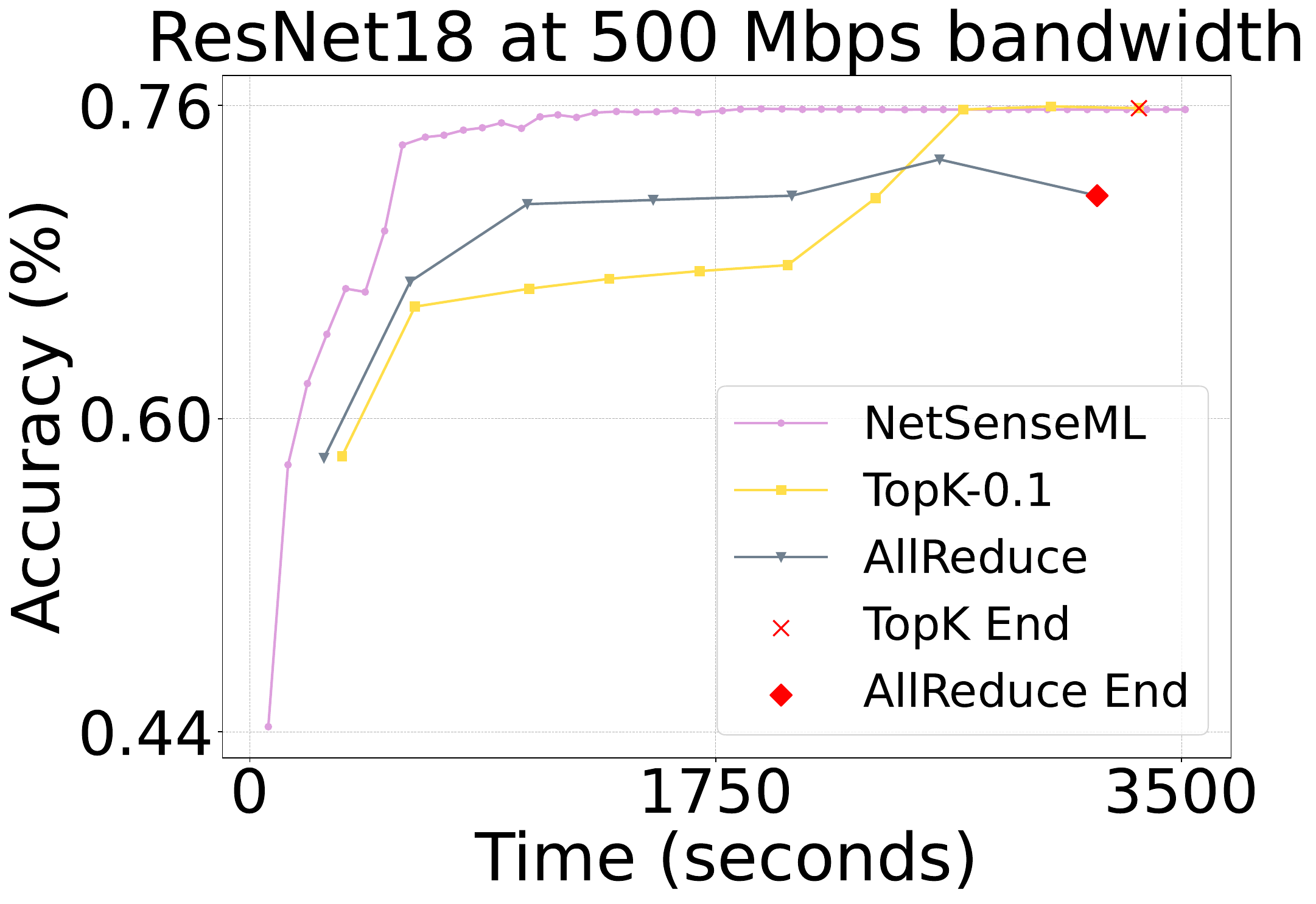} 
  \end{subfigure}
  \begin{subfigure}[b]{0.32\textwidth}
    \centering
    \includegraphics[scale=0.11]{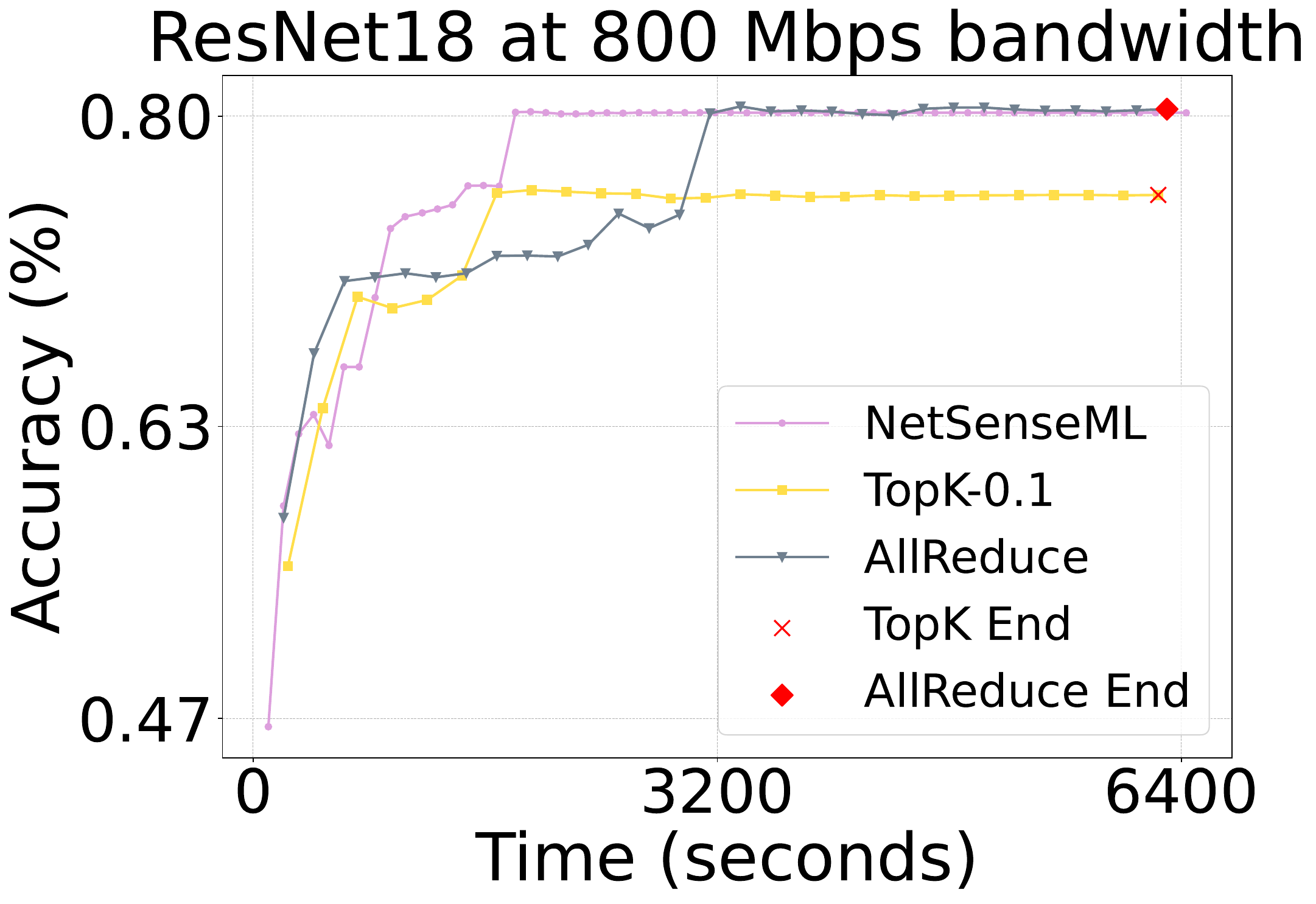} 
  \end{subfigure}
  \caption{Time to accuracy (TTA) comparison under different bottleneck bandwidth conditions with ResNet18}
  \label{fig:tta_comparison_resnet}
\end{figure*}

Fig.~\ref{fig:tta_comparison_resnet} shows that NetSenseML consistently finds an appropriate compression ratio to maintain stable training and converge effectively, even under bandwidth bottlenecks. At 200 Mbps, NetSenseML significantly improves convergence time and stability compared to TopK and AllReduce, which exhibit both slower convergence and reduced stability. NetSenseML maintains a consistent accuracy curve without major fluctuations, while TopK-0.1 suffers from instability. At 500 Mbps, NetSenseML achieves the target accuracy $5 \times$  faster than TopK-0.1. NetSenseML surpasses AllReduce in both convergence speed and stability. Under 800 Mbps, NetSenseML retains its advantage, converging more quickly and steadily to a higher accuracy.

\begin{figure*}[t]
  \centering
  \begin{subfigure}[b]{0.32\textwidth}
    \centering
    \includegraphics[scale=0.11]{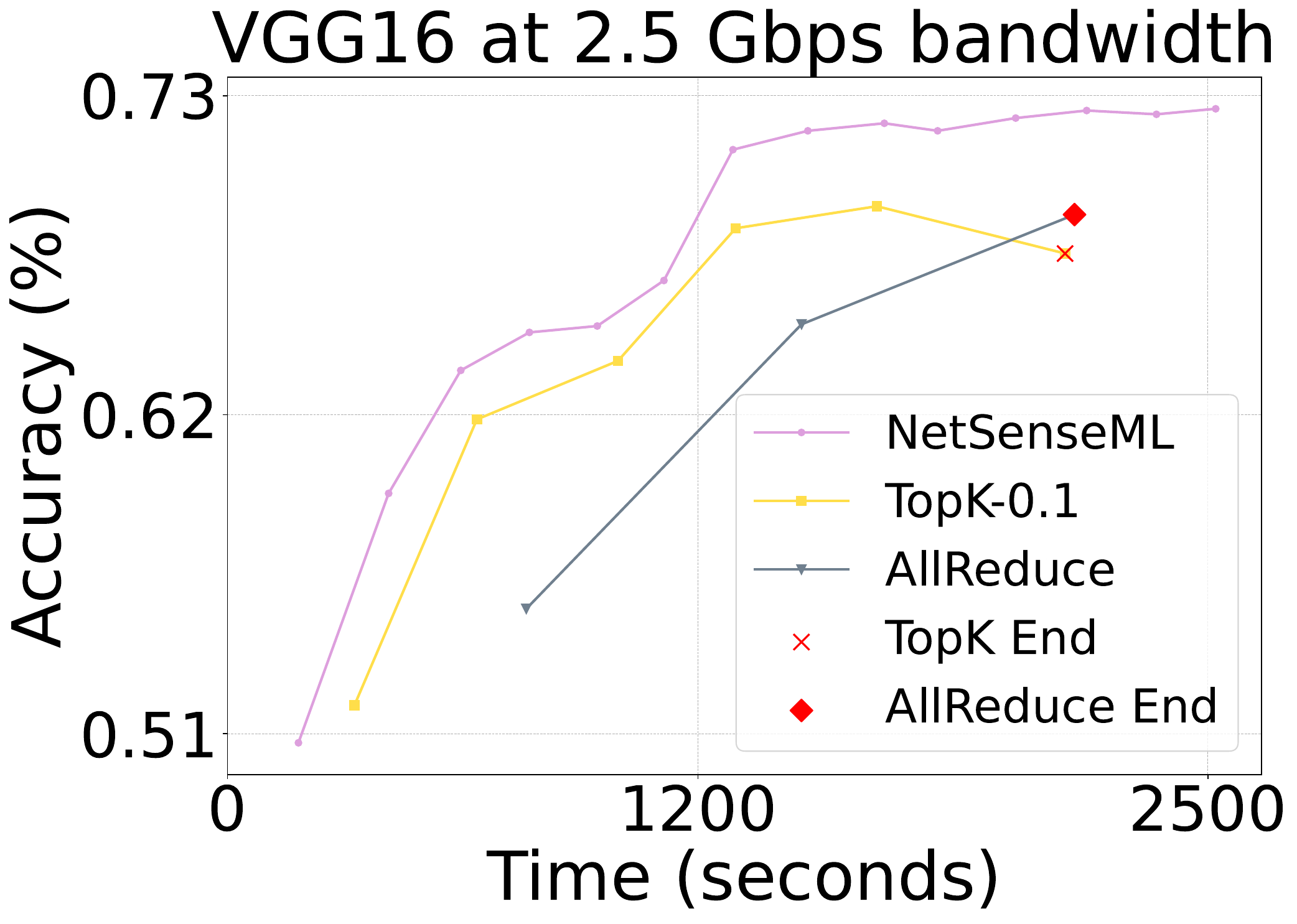}
  \end{subfigure}
  \begin{subfigure}[b]{0.32\textwidth}
    \centering
    \includegraphics[scale=0.11]{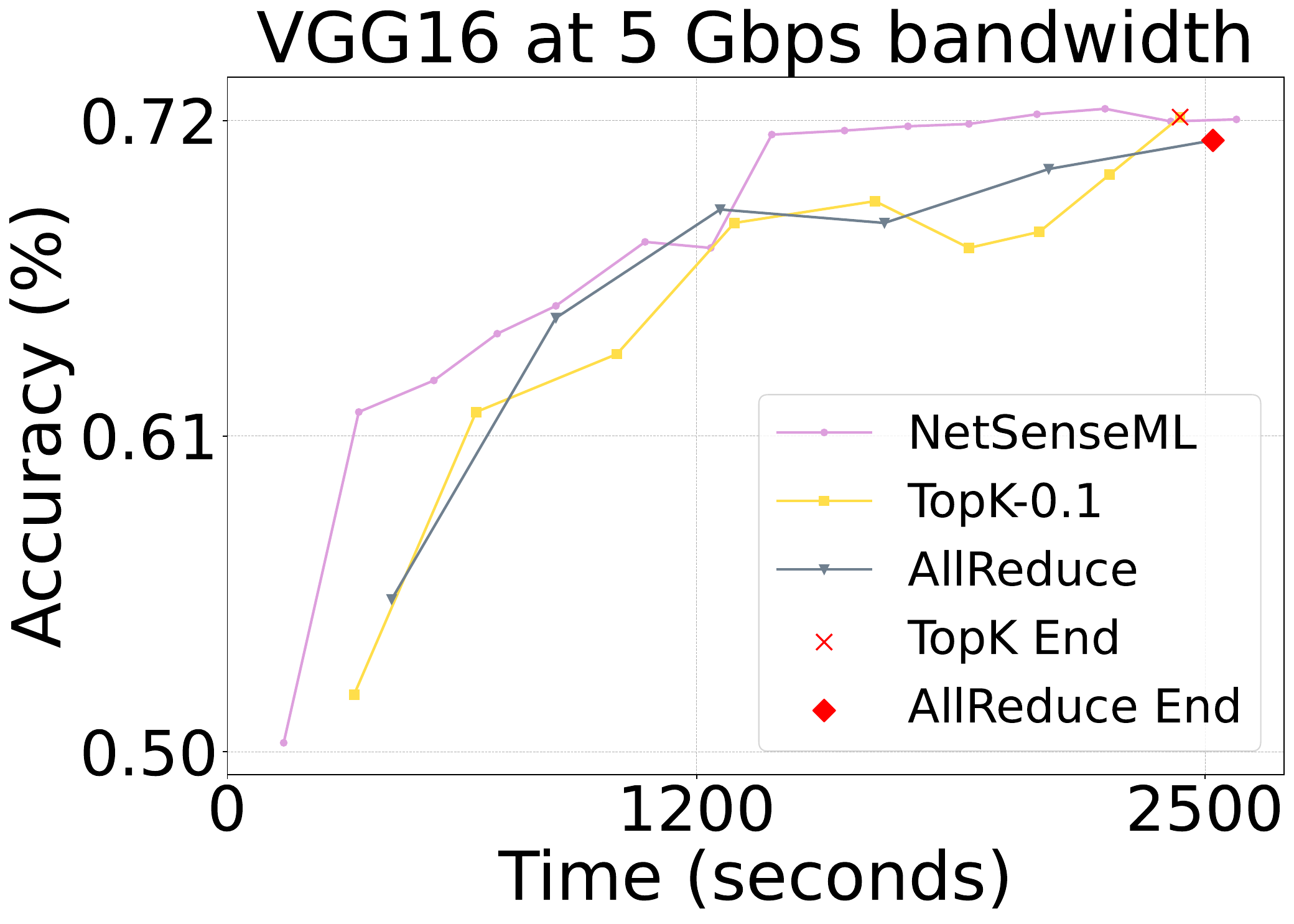}
  \end{subfigure}
  \begin{subfigure}[b]{0.32\textwidth}
    \centering
    \includegraphics[scale=0.11]{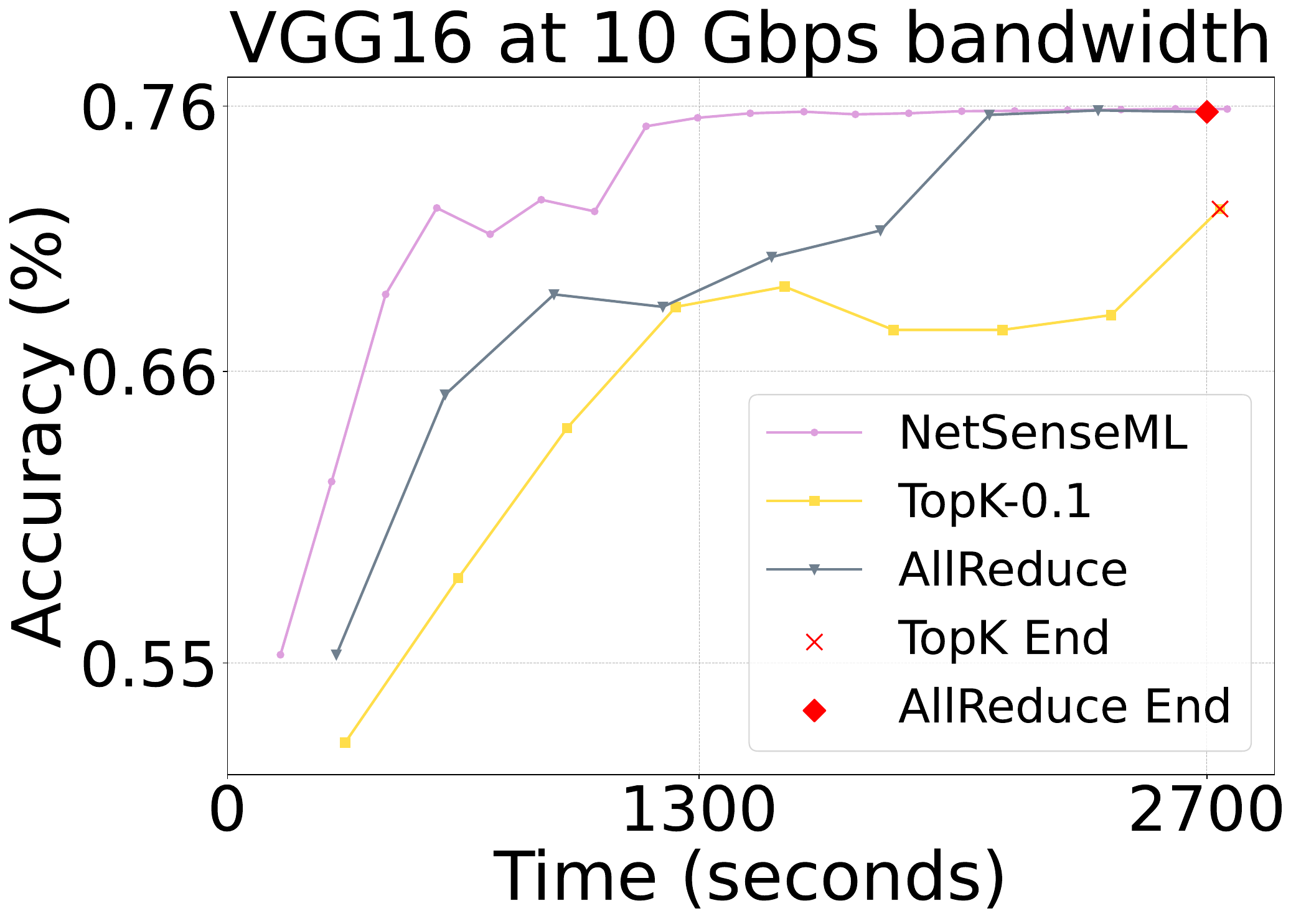}  
  \end{subfigure}
  \caption{Time to accuracy (TTA) comparison under different bottleneck bandwidth conditions with VGG16}
  \label{fig:tta_comparison_vgg}
\end{figure*}

\begin{table*}[t]
\scriptsize 
\centering
\caption{Performance Comparison of ResNet18 under NetSenseML and Other Methods: Training Throughput (samples/second), Convergence Time (seconds), and Best Test Accuracy (\%)}
\label{tab:resnet18}
\begin{tabular}{lcccc}
\toprule
Method & Bottleneck Bandwidth & Test Accuracy & Training Throughput   & Convergence Time \\
\midrule
\rowcolor{gray!20} NetSenseML & 200Mbps & 75.79\% & 642.90 & 1575 \\
AllReduce & 200Mbps & 67.34\% & 42.20 & N/A \\
TopK-0.1 & 200Mbps & 66.52\% & 65.31 & N/A \\
\midrule
\rowcolor{gray!20} NetSenseML & 500Mbps & 76.02\% & 744.24 & 1090 \\
AllReduce & 500Mbps & 73.24\% & 179.85 & N/A \\
TopK-0.1 & 500Mbps & 75.62\% & 144.27 & 3252 \\
\midrule
\rowcolor{gray!20} NetSenseML & 800Mbps & 80.18\% & 824.56 & 1808 \\
AllReduce & 800Mbps & 80.35\% & 283.60 & 3150 \\
TopK-0.1 & 800Mbps & 75.68\% & 234.77 & 1920 \\
\bottomrule
\end{tabular}
\end{table*}

\begin{table*}
\scriptsize 
\centering
\caption{Performance Comparison of VGG16 under NetSenseML and Other Methods}
\label{tab:resnet18}
\begin{tabular}{lcccc}
\toprule
Method & Bottleneck Bandwidth & Test Accuracy & Training Throughput   & Convergence Time \\
\midrule
\rowcolor{gray!20} NetSenseML & 2.5Gbps & 72.04\% & 172.80 & 2520 \\
AllReduce & 2.5Gbps &  66.52\% & 43.20 & N/A \\
TopK-0.1 & 2.5Gbps & 65.12\% & 96.32 & N/A \\
\midrule
\rowcolor{gray!20} NetSenseML & 5Gbps &  72.55\% & 199.46 & 2040 \\
AllReduce & 5Gbps & 71.31\% & 91.58 & N/A \\
TopK-0.1 & 5Gbps & 72.12\% & 128.61 &  2436 \\
\midrule
\rowcolor{gray!20} NetSenseML & 10Gbps & 75.90\% & 340.35 & 1808 \\
AllReduce & 10Gbps & 75.84\% & 129.87 & 2454 \\
TopK-0.1 & 10Gbps & 72.30\% & 148.61 & 2736 \\
\bottomrule
\end{tabular}
\end{table*}


\textbf{Training throughput and communication efficiency.} Table~\ref{tab:resnet18} presents a comparison of training throughput and convergence times for NetSenseML, AllReduce, and TopK-0.1 under different bandwidth conditions. 

The results demonstrate the robustness of NetSenseML in achieving optimal accuracy with significantly reduced convergence time, particularly in challenging low-bandwidth environments.

NetSenseML consistently maintains a specific training throughput across different bandwidth conditions. Taking ResNet18 as an example, when network conditions match the communication scale, avoiding network congestion and preventing increased RTT or packet loss and retransmission, NetSenseML identifies a higher compression rate to avoid compression, optimizing the overall performance. When network limitations become the performance bottleneck for the entire training cluster, such as in cases of network congestion in cloud environments or operational issues in data centers, traditional methods like AllReduce and static compression algorithms like TopK struggle. These methods often slow down the gradient synchronization process, subsequently hindering the entire training process. In contrast, NetSenseML effectively mitigates this issue, adapting dynamically to maintain efficient training.

An interesting observation is that when training ResNet18 under a network bottleneck bandwidth of 200 Mbps, the training throughput of the TopK compression algorithm was higher than that of AllReduce, which aligns with our intuition since TopK has a smaller overall communication volume. However, when the bottleneck bandwidth was increased to 500 Mbps and 800 Mbps, AllReduce achieved higher training throughput compared to TopK-0.1. This can be attributed to the use of the AllGather communication pattern by TopK for gradient synchronization, whereas the NCCL AllReduce operation exhibits a high degree of parallelism, allowing it to efficiently utilize the network link bandwidth.



\begin{figure}[t]
  \centering
  \begin{minipage}{0.45\textwidth}
    \centering
    \includegraphics[scale=0.25]{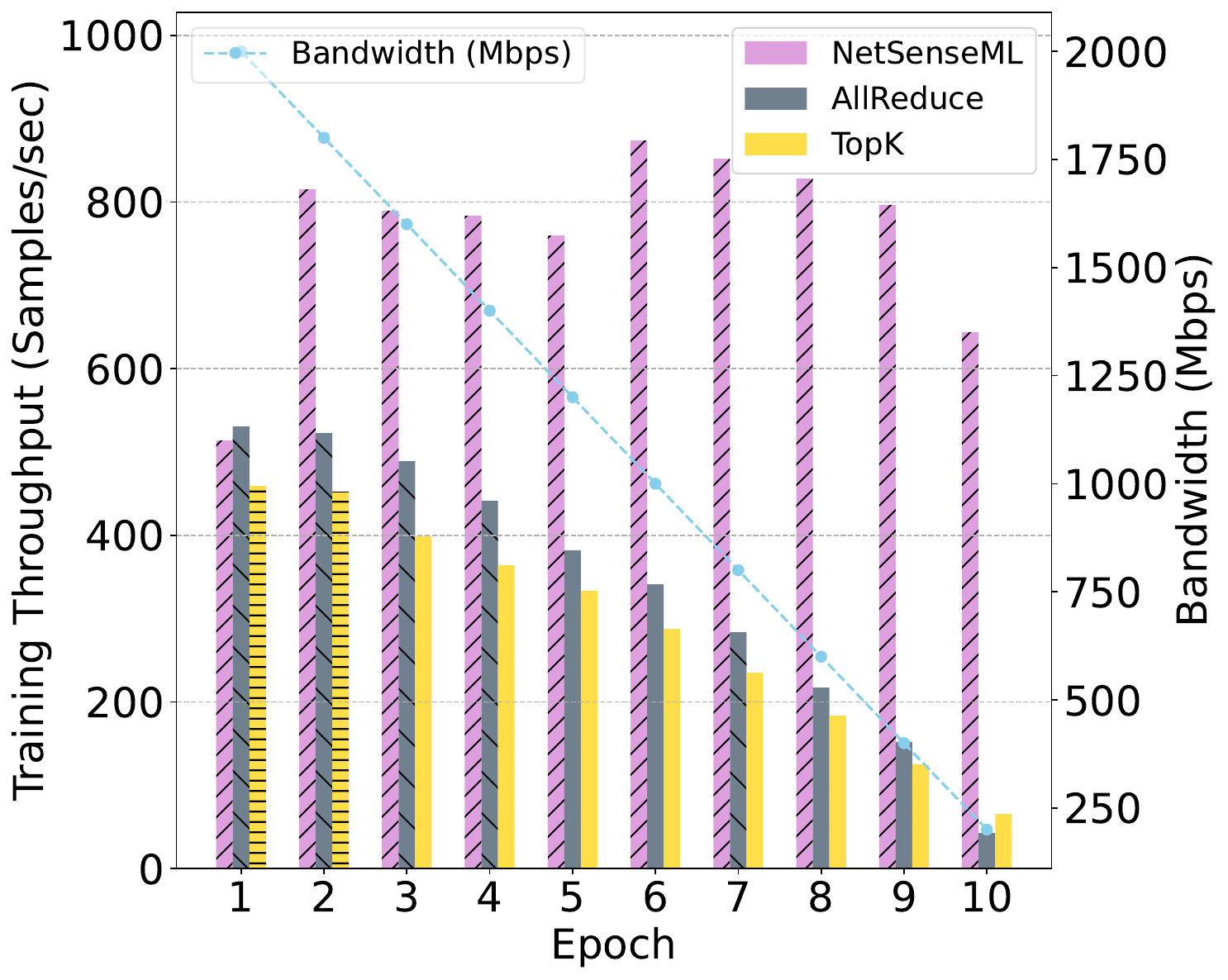}
    \caption{Impact of Bandwidth Degradation on Training Throughput of ResNet18}
    \label{fig:scenario2_resnet18}
  \end{minipage} \hfill
  \begin{minipage}{0.45\textwidth}
    \centering
    \includegraphics[scale=0.25]{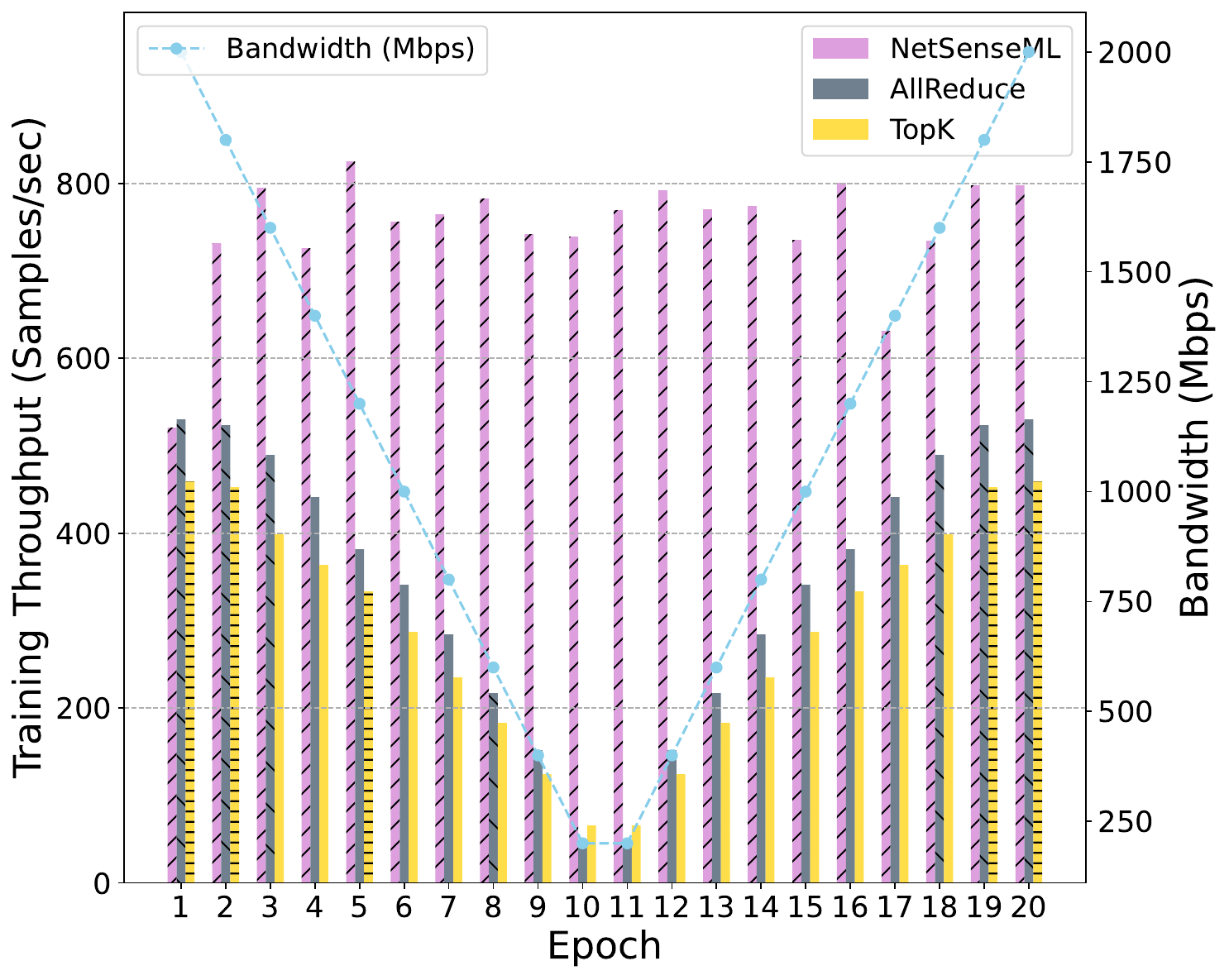}
    \caption{Impact of Bandwidth Fluctuation on Training Throughput of ResNet18}
    \label{fig:scene3}
  \end{minipage}
\end{figure}


\textbf{Dynamic training throughput in degrading network conditions.} The adaptability of NetSenseML to changing bandwidth conditions is illustrated in Scenario 2, where dynamic training performance is compared between different levels of bandwidth. Using the ResNet18 model, we gradually decreased the bottleneck bandwidth - from 2000 to 200 Mbps in steps of 200 Mbps - and measured training performance at each level. This approach highlights the advantage of NetSenseML over TopK and AllReduce. As network conditions deteriorate, NetSenseML is able to maintain higher throughput by reducing gradient transmission, while TopK and AllReduce maintain constant data volumes, leading to network congestion and reduced throughput.

Fig.~\ref{fig:scenario2_resnet18} presents the training throughput of different methods as the initial bottleneck bandwidth of 2000 Mbps gradually decreases to 200 Mbps. In the first epoch, NetSenseML needed time to determine an appropriate compression ratio for the available bandwidth, initially leading to suboptimal throughput. However, as training progressed, NetSenseML adapted effectively, adjusting the compression ratio to accommodate increasingly constrained and adverse network conditions. This adaptability allowed it to maintain consistent training throughput, which is highly valuable for cloud scenarios where programmers need to quickly observe the final convergence of the model. In contrast, TopK and AllReduce were unable to adapt to the decreasing bandwidth, resulting in lower training throughput as network constraints intensified.

\textbf{Dynamic training throughput in Fluctuating network conditions.} In Scenario 3, we used the ResNet18 model, and other configurations were kept the same as in Scenario 2. To illustrate the coexistence of our training task with other network applications, we ran iperf3 \cite{iperf} in a multiprocess setup to compete for the network link, thereby preempting the link between switches and imposing constraints on network traffic during training. Fig.~\ref{fig:scene3} shows that regardless of how dynamically competing trafficies, NetSenseML is consistently able to proceed with stable training throughput, demonstrating a significantly higher level of stability compared to static collective communication methods such as AllReduce and TopK.

\section{Conclusions}
\label{sec:conclusion}

Network fluctuations in bandwidth and latency often degrade resource efficiency, slow model convergence, and reduce accuracy, especially in heterogeneous environments. To address this, we propose NetSenseML, which is an adaptive compression algorithm that dynamically adjusts gradient compression ratios based on real-time network performance.

By monitoring network conditions and estimating bandwidth, NetSenseML applies quantization, pruning, and sparsification to optimize both critical gradient retention and model generalizability. NetSenseML performs well in adapting to a variety of network conditions, ensuring robust and efficient gradient transmission. This adaptability significantly improves the performance of model training and convergence rates in various scenarios, effectively mitigating the typical trade-offs between transmission efficiency, convergence speed, and accuracy.

\section{Acknowledgements}

Guangdong provincial project 2023QN10X048, Guangzhou Municipal Key Laboratory on Future Networked Systems (2024A03J0623),
the Guangzhou Municipal Science and Technology Project 2023A03J0011, the Guangdong provincial project 2023ZT10X009, and the
Natural Science Foundation of China (U23A20339).

Yisu Wang was responsible for the design of the experimental methodology and conducting the experiments. Xinjiao led the writing and refinement of the experimental section. They should be considered co-first authors.

\subsubsection*{Disclosure of Interests}
The authors have no competing interests to declare that are relevant to the content of this article.

\bibliographystyle{splncs04}
\bibliography{ref}






\end{document}